\documentclass[runningheads,a4paper]{llncs}

\usepackage{amssymb}
\usepackage{amsmath}
\usepackage{graphicx}
\usepackage{subfigure}
\usepackage{url}
\usepackage{algorithm}
\usepackage{algorithmic}
\usepackage{multirow}
\usepackage{rotating}
\usepackage[table]{xcolor}
\usepackage{color}      
\usepackage{float}

\DeclareFixedFont{\afacc}{OT1}{phv}{m}{n}{10}

\begin{document}

\mainmatter  

\title{Ordinal Rating of Network Performance and Inference by Matrix Completion}


%
%


\author{Wei Du\inst{1}, Yongjun Liao\inst{1}%
, Pierre Geurts\inst{2}, Guy Leduc\inst{1}}
\authorrunning{Du et al.}

\institute{Research Unit in Networking (RUN), University of Li\`ege, Belgium\\
\email{weidu@montefiore.ulg.ac.be,\{yongjun.liao,guy.leduc\}@ulg.ac.be}
\and Systems and Modeling, University of Li\`ege, Belgium\\
\email{p.geurts@ulg.ac.be}
}

%

\maketitle

\begin{abstract}
This paper addresses the large-scale acquisition of end-to-end network performance. We made two distinct contributions: ordinal rating of network performance and inference by matrix completion. The former reduces measurement costs and unifies various metrics  which eases their processing in applications. The latter enables scalable and accurate inference with no requirement of structural information of the network nor geometric constraints. By combining both, the acquisition problem bears strong similarities to recommender systems. This paper investigates the applicability of various matrix factorization models used in recommender systems. We found that the simple regularized matrix factorization is not only practical but also produces accurate results that are beneficial for peer selection.

\end{abstract}

\section{Introduction}
\label{sec:introduction}

The knowledge of end-to-end network performance is beneficial to many Internet applications~\cite{crovella_IM}. 
To acquire such knowledge, there are two main challenges. First, the performance of a network path can be characterized by various metrics which differ largely. On the one hand, the wide variety of these metrics renders their processing difficult in applications. On the other hand, although having been studied for decades, network measurement for many metrics still suffers from high costs and low accuracies. Second, it is critical to efficiently monitor the performance of the entire network. As the number of network paths grows quadratically with respect to the number of network nodes, active probing of all paths on large networks is clearly infeasible.

In this paper, we address these challenges by two distinct contributions: ordinal rating of network performance and inference by matrix completion.

{\setlength{\parindent}{0pt}\setlength{\parskip}{1ex}\textbf{Ordinal Rating of Network Performance.} Instead of quantifying the performance of a network path by the exact value of some metric, we investigate the rating of network performance by ordinal numbers of $1,2,3,\ldots$, with larger value indicating better performance, regardless of the metric used. For the following reasons, ordinal ratings are advantageous over exact metric values.}
\begin{itemize}
 \item Ratings carry sufficient information that already fulfills the requirements of many Internet applications. For example, streaming media cares more about whether the available bandwidth of a path is high enough to provide smooth playback quality. In peer-to-peer applications, although finding the nearest nodes to communicate with is preferable, it is often enough to access nearby nodes with limited loss compared to the nearest ones. Such objective of finding ``good-enough'' paths can be well served using the rating information.
 \item Ratings are coarse measures that are cheaper to obtain. They are also stable and better reflect long-term characteristics of network paths, which means that they can be probed less often.
 \item The representation by ordinal numbers not only allows the rating information to be encoded in a few bits, saving storage and transmission costs, but also unifies various metrics and eases their processing in applications.
\end{itemize}

{\setlength{\parindent}{0pt}\setlength{\parskip}{1ex}\textbf{Inference by Matrix Completion.} We then address the scalability issue by network inference that measures a few paths and predicts the performance of the other paths where no direct measurements are made. In particular, we formulate the inference problem as matrix completion where a partially observed matrix is to be completed~\cite{candes2010}. Here, the matrix contains performance measures between network nodes with some of them known and the others unknown and thus to be filled. Comparing to previous approaches \cite{dabek04,Chen:2004,kriging2006}, our matrix completion formulation relies on neither structural information of the network nor geometric constraints. Instead, it exploits the spatial correlations across network measurements, which have long been observed in various research~\cite{Chen:2004,kriging2006}.}

{\setlength{\parskip}{1ex} By integrating the ordinal rating of network performance and the matrix completion formulation, a great benefit is that the acquisition problem bears strong similarities to recommender systems which have been well studied in machine learning. Thus, a particular focus of this paper is to investigate the applicability of various recommender system solutions to our network inference problem. In particular, we found that the simple regularized matrix factorization is not only practical but also produces accurate results that are beneficial for the application on peer selection.}

Previous work on network inference focused on predicting the metric values of network paths. For example, \cite{Chen:2004,kriging2006} solved the inference problem by using the routing table of the network. In contrast, Vivaldi~\cite{dabek04} and DMFSGD~\cite{liao_dmf2012} built the inference models, without using network topology information, based on Euclidean embedding and on matrix completion. The same DMFSGD algorithm was adapted in \cite{liao_conext2011} to classify network performance into binary classes of either ``good'' or ``bad''. Based on \cite{liao_conext2011}, this paper goes further and studies ordinal ratings of network performance and their inference by solutions to recommender systems.

The rest of the paper is organized as follows. Section~\ref{sec:performance} describes the metrics and the rating of network performance. Section~\ref{sec:inference} introduces network inference by matrix completion. Section~\ref{sec:experiment} presents the experimental results and the application on peer selection. Section~\ref{sec:conclusion} gives conclusions.

\section{Network Performance}
\label{sec:performance}

\subsection{Metrics}

End-to-end network performance is a key concept at the heart of networking~\cite{crovella_IM}. 
Numerous metrics have been designed to serve various objectives. For example, delay-related metrics measure the response time between network nodes and are interested by downloading services. Bandwidth-related metrics indicate the data transmission rate over network paths and are concerned by online streaming. On the acquisition of these metrics, great efforts have been made and led to various measurement tools. However, the acquisition for some metrics still suffers from high costs and low accuracies. For example, measuring the available bandwidth of a path often requires to congest the path being probed. 

In this paper, we focus on two commonly-used performance metrics, namely round-trip time (RTT) and available bandwidth (ABW).

\subsection{Ordinal Rating}

Acquiring end-to-end network performance amounts to determining some quantity of a chosen metric. As mentioned earlier, ratings go beyond exact values in a number of ways. In addition, ratings reflect better the experience of end users to the Quality of Service (QoS), by which network performance should be defined.

Ratings take ordinal numbers in the range of $[1,R]$, where $R=5$ in this paper. The different levels of rating indicate qualitatively how well network paths would perform, i.e., $1$--``very poor'', $2$--``poor'', $3$--``ordinary'', $4$--``good'' and $5$--``very good''.

\begin{figure}[b]
\centering
\subfigure{\includegraphics[width=0.4\columnwidth]{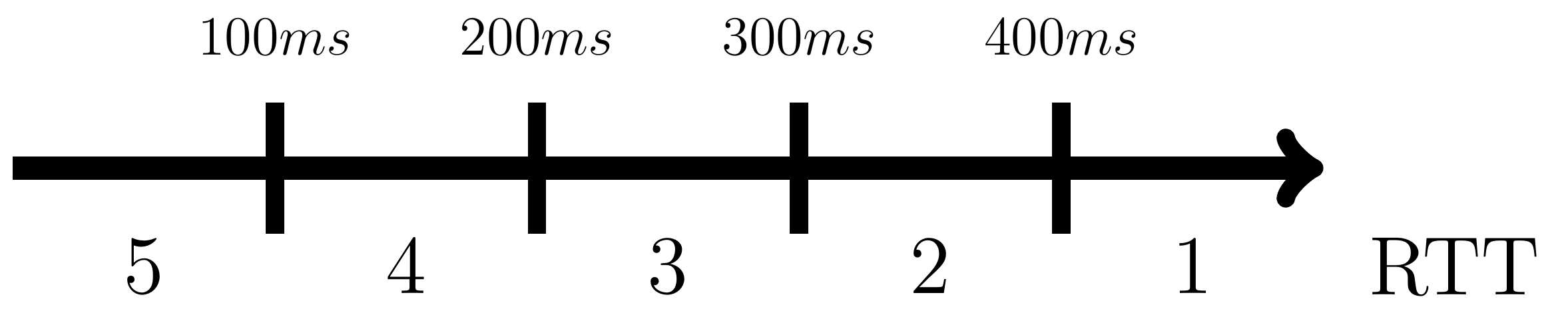}}\qquad \qquad
\subfigure{\includegraphics[width=0.4\columnwidth]{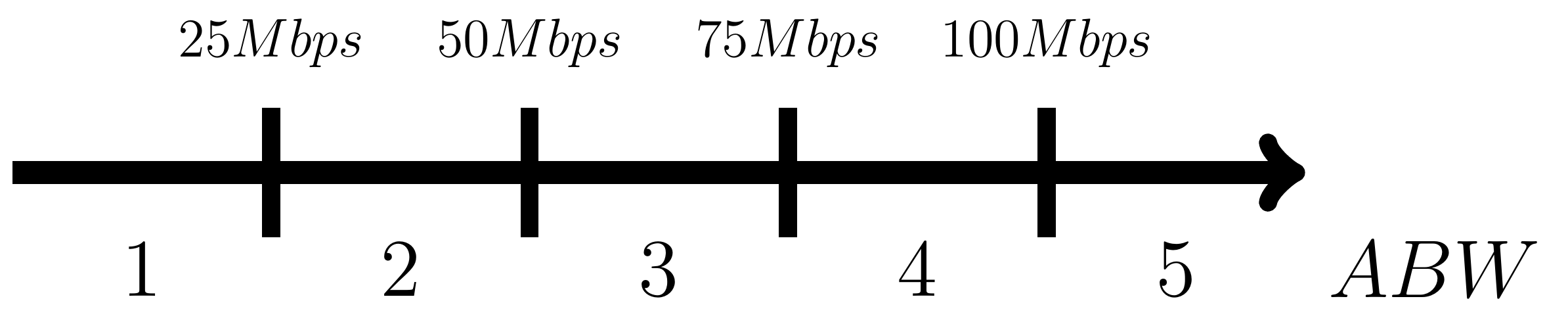}}
\caption{Examples of quantification of metric values into ratings of $[1,5]$. }
\label{fig:rating}
\end{figure} 

Generally, ratings can be acquired by vector quantization that partitions the range of the metric into $R$ bins using $R-1$ thresholds, $\tau=\{\tau_1,\ldots,\tau_{R-1}\}$, and determines to which bins metric values belong, as illustrated in Figure~\ref{fig:rating}. The thresholds can be chosen evenly or unevenly according to the requirements of the applications. Clearly, rating a path is cheaper than measuring the exact value as we only need to determine if the value is within a certain range defined by the thresholds. This holds for most, if not all, metrics, since data acquisition generally undergoes the accuracy-versus-cost dilemma that accuracy always comes at a cost. The cost reduction is particularly significant for ABW.

\subsection{Intelligent Peer Selection}
 
For many Internet applications, the goal of acquiring end-to-end network performance is to achieve the QoS objectives for end users. Examples include choosing low-delay peers to communicate with in overlay networks or choosing a high-bandwidth mirror site from which to stream media. In these examples, intelligent peer selection is desired to optimize services by finding for each node a peer that is likely to respond fast and well. 
 
The question is, to achieve this goal, should we use metric values or ratings of network paths? On the one hand, the knowledge of metric values allows to find the globally optimal node over the entire network. On the other hand, although the rating information only allows to find ``good-enough'' node, ratings are cheaper to obtain. Thus, it is interesting to study the optimality of peer selection based on ratings and on metric values.

\section{Inference by Matrix Completion}
\label{sec:inference}

\subsection{Fundamentals of Matrix Completion}
\label{sec:mc}
Matrix completion addresses the problem of recovering a low-rank matrix from a subset of its entries. In practice, a real data matrix $X$ is often full rank but with a rank $r$ dominant components. That is, $X$ has only $r$ significant singular values and the others are negligible. In such case, a matrix of rank $r$, denoted by $\hat{X}$, can be found that approximates $X$ with high accuracy~\cite{candes2010}.
 
Generally, matrix completion is solved by the low-rank approximation,
\begin{equation}
\label{eq:lossfunc_rankmf}
P_\varOmega(X) \approx P_\varOmega(\hat{X})\ \text{s.t. \textit{Rank}}(\hat{X})\leqslant r.
\end{equation}
$\varOmega$ is the set of observed entries and $P_\varOmega$ is a sampling function that preserves the entries in $\varOmega$ and turns the others into $0$. In words, we try to find a low-rank matrix $\hat{X}$ that best approximates $X$ for the observed entries.

The rank function is difficult to optimize or constrain, since it is neither convex nor continuous. Alternatively, the low-rank constraint can be tackled directly by adopting some compact representation. For example, as $\mathrm{rank}(\hat{X})\leqslant r$,
\begin{equation}
\label{eq:mc_compact}
\hat{X}=UV^T,
\end{equation}
where $U$ and $V$ are matrices of $n\times r$. Thus, we can look for the pair $(U,V)$, instead of $\hat{X}$, such that 
\begin{equation}
\label{eq:mc_lowrankUV}
P_\varOmega(X)\approx P_\varOmega(UV^T).
\end{equation}
The class of techniques to solve eq.~\ref{eq:mc_lowrankUV} is generally called matrix factorization (MF). As the pair $(U,V)$ has $2nr$ entries in contrast to the $n^2$ for $\hat{X}$, matrix factorization is much more appealing for large matrices.


\subsection{Network Inference}
The network performance inference is formulated as a matrix completion problem. In this context, $X$ is a $n\times n$ matrix constructed from a network of $n$ nodes. The entry $x_{ij}$ is some performance measure, a rating in our case, of the path from node $i$ to node $j$. $X$ is largely incomplete as many paths are unmeasured.

\begin{figure}[t]
\centering
\includegraphics[width=0.35\columnwidth]{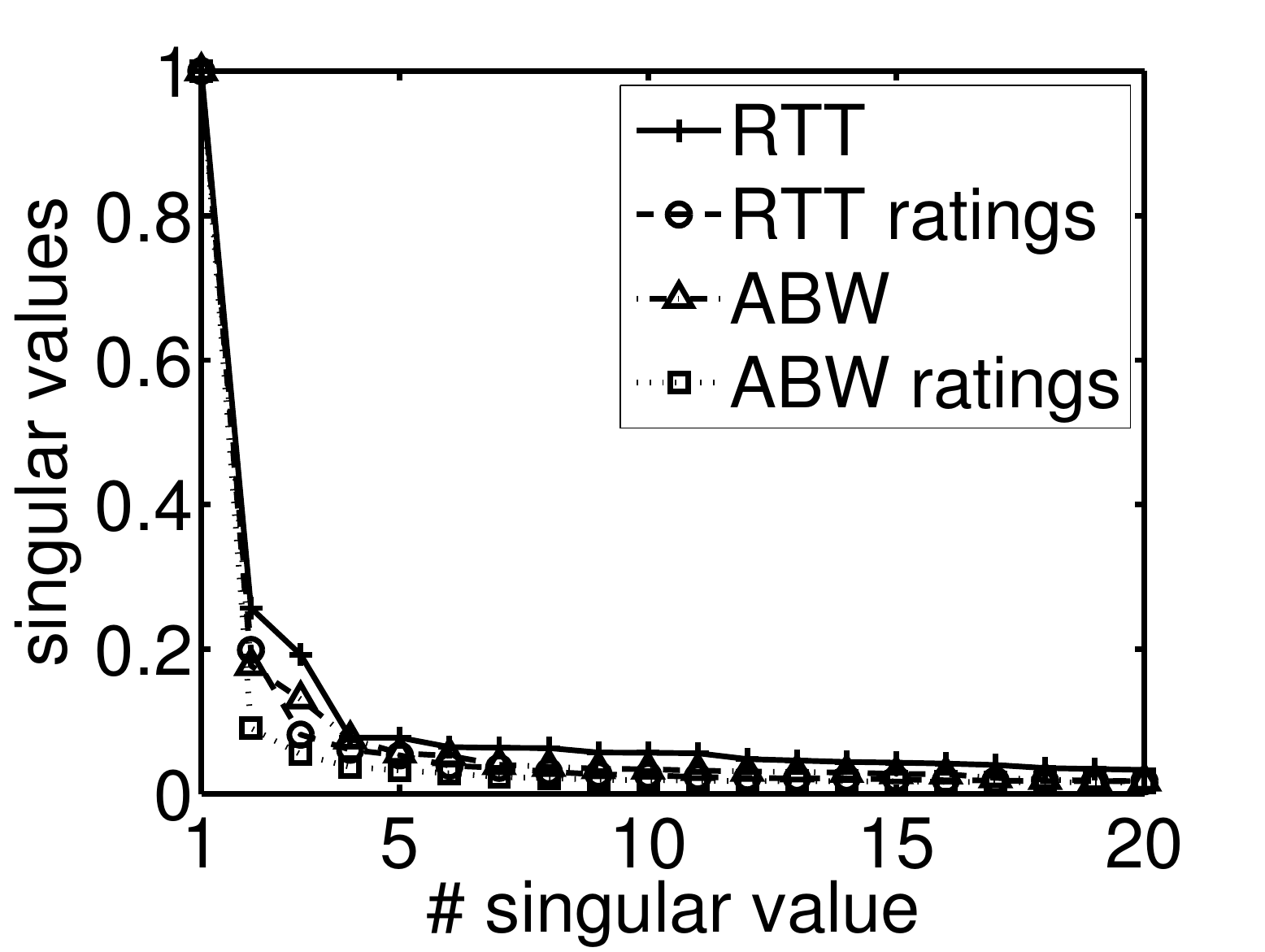}
\caption{\small{The singular values of a 
RTT matrix and a 
ABW matrix, and of the corresponding rating matrices. The singular values are normalized by the largest one.}}
\label{fig:meridian2255_sv}
\end{figure}

Network inference is feasible because network measurements are correlated across paths. The correlations come largely from link sharing between network paths~\cite{Chen:2004,kriging2006}, due to the topology simplicity in the Internet core where network paths overlap heavily. These correlations induce the related performance matrix to be approximately low-rank and enable the inference problem to be solved by matrix completion. We empirically evaluate the low-rank characteristics of a RTT and a ABW matrix by the spectral plots in Figure~\ref{fig:meridian2255_sv}. It can be seen that the singular values of both the original matrices and of the related rating matrices decrease fast. This low-rank phenomenon has been consistently observed in many research \cite{Chen:2004,kriging2006,liao_dmf2012,liao_conext2011}.

As performance measures are ratings, the inference problem bears strong similarities to recommender systems which predict the rating that a user would give to an item such as music, books, or movies~\cite{Koren09}. In this context, network nodes are users and they treat other nodes as items. In a sense, a rating is a preference measure of how a node would like to contact another node.

A big motivation of recommender systems was the Netflix prize which was given to the BellKor's Pragmatic Chaos team in 2009~\cite{netflixprize}. In the sequel, the prize-winning solution is called BPC. BPC integrated two classes of techniques based on neighborhood models and on matrix factorization. Neighborhood models exploit the similarities between users and between items. Calculating similarities requires a sufficient number of ratings which may not be available in our problem. Thus, we focus in this paper on the applicability of matrix factorization and leave the study on neighborhood models as future work.

\subsection{Matrix Factorization}

The goal of MF is to find $U$ and $V$ such that $UV^T$ is close to $X$ at the observed entries in $\varOmega$. This section discusses various MF models that were integrated in BPC including RMF, MMMF and NMF~\cite{netflixprize}.

{\setlength{\parindent}{0pt}\setlength{\parskip}{1ex}\textbf{RMF}
Regularized matrix factorization (RMF)~\cite{Koren09} adopts the widely-used $L_2$ loss function and solves 
\begin{equation}
 \label{eq:mc_rmf}
 \mathrm{min}\sum_{ij\in\varOmega}(x_{ij}-u_iv_j^T)^2+\lambda\sum_{i=1}^n(u_iu_i^T+v_iv_i^T),
\end{equation}
where $u_i$ and $v_i$ are row vectors of $U$ and $V$ and $x_{ij}$ is the $ij$th entry of $X$. The second term is the regularization which restricts the norm of $U$ and $V$ so as to prevent overfitting. $\lambda$ is the regularization coefficient.} 

The unknown entries in $X$ are predicted by
\vspace{-.15cm}
\begin{equation}
 \label{eq:mc_prediction}
 \hat{x}_{ij}=u_iv_j^T,\ \mathrm{for}\ ij\notin\varOmega.
\end{equation}
Note that $\hat{x}_{ij}$ is real-valued and has to be rounded to the closest integer in the range of $[1,R]$.

{\setlength{\parindent}{0pt}\setlength{\parskip}{1ex}\textbf{MMMF}
Max-margin matrix factorization (MMMF) solves the inference problem by ordinal regression~\cite{FMMMF}. As RMF, the unknown entries in $X$ are predicted by eq.~\ref{eq:mc_prediction}. However, instead of rounding, the real-valued estimate $\hat{x}_{ij}$ is related to the ordinal rating $x_{ij}$ by using $R-1$ thresholds $\theta_1, \ldots, \theta_{R-1}$ and requiring
\vspace{-.15cm}
\begin{equation}
\label{eq:mmmf_constraint}
\theta_{x_{ij}-1}<\hat{x}_{ij}=u_iv_j^T<\theta_{x_{ij}}, 
\end{equation}
where for simplicity of notation $\theta_0=-\infty$ and $\theta_R=\infty$. In words, the value of $\hat{x}_{ij}$ does not matter, as long as it falls in the range of $[\theta_{r-1},\theta_r]$ for $x_{ij}=r$.}

In practice, it is impossible to have eq.~\ref{eq:mmmf_constraint} satisfied for every $x_{ij}$. Thus, we penalize the violation of the constraint and solve 
\vspace{-.2cm}
\begin{equation}
\label{eq:mc_mmmf}
\mathrm{min}\sum_{ij\in\varOmega}\sum_{r=1}^Rl(T_{ij}^r,\theta_r-u_iv_j^T)+\lambda\sum_{i=1}^n(u_iu_i^T+v_iv_i^T),
\end{equation}
where $T_{ij}^r=1$ if $x_{ij} \leqslant r$ and $-1$ otherwise. Essentially, eq.~\ref{eq:mc_mmmf} consists of a number of binary classifications each of which compares an estimate $\hat{x}_{ij}$ with a threshold. The loss function $l$ can be any classification loss function, among which the smooth hinge loss function is used.

{\setlength{\parindent}{0pt}\setlength{\parskip}{1ex}\textbf{NMF}
Non-negative matrix factorization (NMF)~\cite{Lee01algorithmsfor} incorporates an additional constraint that all entries in $U$ and $V$ have to be non-negative so as to ensure the non-negativity of $\hat{X}$.}

Besides, NMF often uses the divergence to measure the difference between $X$ and $\hat{X}$, defined as
\vspace{-.2cm}
\begin{equation}
 D(X||\hat{X})=\sum_{ij\in\varOmega}(x_{ij}\mathrm{log}\frac{x_{ij}}{\hat{x}_{ij}}-x_{ij}+\hat{x}_{ij}).
\end{equation}
Thus, NMF solves
\vspace{-.2cm}
\begin{align}
 \label{eq:mc_nmf}
 \mathrm{min}\ & D(X||UV^T)+\lambda\sum_{i=1}^n(u_iu_i^T+v_iv_i^T).
&\mathrm{s.t.}\ U\geqslant 0,\ V\geqslant 0\nonumber
\end{align}
As RMF, $\hat{x}_{ij}$ is real-valued and has to be rounded to the closest integer in the range of $[1,R]$.

{\setlength{\parindent}{0pt}\setlength{\parskip}{1ex}\textbf{MF ENSEMBLES}
The success of BPC built on the idea of the ensemble which learns multiple models and combines their outputs for prediction~\cite{netflixprize}. 
In machine learning, usually several different models can give similar accuracy on the training data but perform unevenly on the unseen data. In this case, a simple vote or average of the outputs of these models can reduce the variance of the predictions. In this paper, we combine the above RMF, MMMF and NMF. The final prediction result is the average of the predictions by different MF models.}

%


\subsection{Implementation Details}

{\setlength{\parindent}{0pt}\setlength{\parskip}{1ex}\textbf{Inference By Stochastic Gradient Descent}
In BPC, different MF models are solved by different optimization schemes, some of which are not appropriate for network applications where decentralized processing of data is necessary. In this paper, we adopted Stochastic Gradient Descent (SGD) for all MF models. In short, at each iteration, we pick $x_{ij}$ in $\varOmega$ randomly and update $u_i$ and $v_j$ by gradient descent to reduce the difference between $x_{ij}$ and $u_iv_j^T$. SGD is suitable for network inference, because measurements can be acquired on demand and processed locally at each node. We refer the interested readers to \cite{liao_dmf2012} for the details of the decentralized inference by SGD.}

{\setlength{\parindent}{0pt}\setlength{\parskip}{1ex}\textbf{Neighbor Selection}
In recommender systems, users rate items voluntarily. This is different in network inference where we do have control over data acquisition, i.e., network nodes can actively choose to measure some paths connected to them. Thus, we adopt the system architecture in \cite{dabek04,liao_dmf2012,liao_conext2011} that each node randomly selects $k$ nodes to probe, called neighbors in the sequel.}

Thus, each node collects $k$ ratings from the paths connecting to its neighbors and infers the other unmeasured paths. $k$ has to be chosen by trading off between accuracies and overheads. On the one hand, increasing $k$ always improves accuracies as we measure more and infer less. On the other hand, the more we measure, the higher the overhead is. Thus, as is in~\cite{liao_conext2011}, we set $k=32$ for networks of a few thousand nodes and $k=10$ for a few hundred nodes, leading to about $1-5\%$ available measurements.

{\setlength{\parindent}{0pt}\setlength{\parskip}{1ex}\textbf{Rank $r$}
The most important parameter in MF is the rank $r$. If a given $X$, constructed from a network, is complete, the proper rank can be studied by the spectral plot as Figure~\ref{fig:meridian2255_sv}, under a given accuracy requirement. When $X$ is incomplete, we can only search for the optimal $r$ empirically. On the one hand, $r$ has to be large enough so that enough information in $X$ is kept. On the other hand, a higher-rank matrix has less redundancies and requires more data to recover, increasing measurement overheads. Thus, we choose a small value of $r=10$ as we have only a limited number of measurements.}



\section{Experiments and Evaluations}
\label{sec:experiment}
The evaluations were performed on three datasets including Harvard, Meridian and HP-S3. Among them, Harvard contains dynamic RTT measurements collected from a network of 226 nodes in 4 hours, Meridian is a static RTT matrix of $2500\times 2500$ and HP-S3 is a static ABW matrix of $231\times 231$. More details about these datasets can be found in~\cite{liao_conext2011}. We adopted the common evaluation criterion used for recommender systems, Rooted Mean Square Error (RMSE), 
\vspace{-.1cm}
\begin{equation}
 RMSE = \sqrt{\frac{\sum_{i=1}^n(x_i-\hat{x}_i)^2}{n}}.
\end{equation}
Note that the smaller RMSE is, the better.

\subsection{Obtaining Ratings}
We first obtain ratings from the three datasets. To this end, we partition the range of the metric by the rating threshold $\tau=\{\tau_1,\ldots,\tau_{4}\}$. $\tau$ is set by two strategies: 1. set $\tau$ by the $[20\%,40\%,60\%,80\%]$ percentiles of each dataset; 2. partition evenly the range between $0$ and a large value selected for each dataset.  

Thus, for Strategy 1, $\tau=[48.8,92.2,177.2,280.3]$ms for Harvard, $\tau=[31.6,\\47.3,68.6,97.9]$ms for Meridian, and $\tau=[12.7,34.5,48.8,77.9]$Mbps for HP-S3. For Strategy 2, $\tau=[75,150,225,300]$ms for Harvard , $\tau=[25,50,75,100]$ms for Meridian, and $\tau=[20,40,60,80]$ Mbps for HP-S3. Strategy 2 produces quite unbalanced portions of ratings on each dataset. 

\subsection{Comparison of Different MF Models}
We solved RMF, MMMF and NMF by SGD. The learning rate of SGD $\eta$ equals $0.05$, the regularization coefficient $\lambda$ equals $0.1$, and the rank $r$ is $10$ for all datasets. The neighbor number $k$ is $10$ for Harvard and HP-S3 and $32$ for Meridian. We do not fine tune the parameters for each dataset and for each model, as it is impossible for the decentralized processing. Empirically, MF is not very sensitive to the parameters as the inputs are ordinal numbers of $[1,5]$, regardless of the actual metric and values. For MF ensembles, we generate for each MF model several predictors using different parameters~\cite{netflixprize}. Although maintaining multiple predictors in parallel is impractical, MF ensembles produce the (nearly) optimal accuracy that could be achieved based on MF in a centralized manner.

Table~\ref{tab:rmse1} and \ref{tab:rmse2} show the RMSE achieved using different MF models and different $\tau$-setting strategies. Particularly, we made the following observations. First, RMF generally performs better than MMMF and NMF, MF ensembles perform the best. Second, the improvement of MF ensembles over RMF is only marginal, which is not considered worth the extra cost. Third, the accuracy on Harvard is the worst, which is probably because the dynamic measurements in Harvard were obtained passively, i.e., there was no control over when and which neighbors a node probed. Last, it is clear that different settings of $\tau$ have some impacts to the accuracy, which need to be further studied.Nevertheless, we adopt Strategy 1 by default in the sequel.

\begin{table}[b]
\begin{minipage}{.48\linewidth}
  \centering
    \begin{tabular}{|c|c|c|c|}
\hline
$\tau$ & Harvard & Meridian & HP-S3 \\ \hline
    RMF & 0.9340 & 0.8306 & 0.6754\\ \hline
    MMMF & 0.9688 & 0.8634 & 0.6862\\ \hline
    NMF & 0.9772 & 0.9042 & 0.6820\\ \hline
    MF Ensembles & 0.9205 & 0.8214 & 0.6611\\ \hline
  \end{tabular} 
  \caption{RMSE with $\tau$ set by strategy 1}
  \label{tab:rmse1}
\end{minipage} \hspace{1.5mm}
\begin{minipage}{.48\linewidth}
  \centering
\begin{tabular}{|c|c|c|c|}
\hline
$\tau$ & Harvard & Meridian & HP-S3 \\ \hline
    RMF & 0.9198 & 0.7761 & 0.6669\\ \hline
    MMMF & 0.9193 & 0.8099 & 0.6697\\ \hline
    NMF & 0.9316 & 0.8286 & 0.6742\\ \hline
    MF Ensembles & 0.9043 & 0.7658 & 0.6527\\ \hline
  \end{tabular} 
  \caption{RMSE with $\tau$ set by strategy 2}
    \label{tab:rmse2}
\end{minipage}
\end{table}

We would like to mention that for the Netflix dataset, the RMSE achieved by the Netflix's algorithm cinematch is $0.9525$ and that by BPC is $0.8567$~\cite{netflixprize}. This shows that in practice, the prediction with an accuracy of the RMSE less than 1 is already usable by applications. Thus, by trading off between the accuracy and the practicability, the RMF model is adopted by default in our system. Table~\ref{tab:confusion} shows the confusion matrices achieved by RMF on the three datasets. In these matrices, each column represents the predicted ratings, while each row represents the actual ratings. Thus, the off-diagonal entries represent ``confusions'' between two ratings. It can be seen that while there are mis-ratings, few have an error of $|x_{ij}-\hat{x}_{ij}|>1$, which means that the mis-ratings are under control.
\begin{table}[b]
\centering
\begin{tabular}{|c|c|c|c|c|c|}
\hline
&{1}&{2}&{3}&{4}&{5}\\\hline
{1}&0.68&0.28&0.02&0.01&0.00\\\hline
{2}&0.18&0.60&0.20&0.01&0.00\\\hline
{3}&0.03&0.13&0.66&0.17&0.00\\\hline
{4}&0.03&0.04&0.16&0.67&0.10\\\hline
{5}&0.04&0.03&0.05&0.43&0.46\\\hline
\end{tabular} \hspace{.1cm}
\begin{tabular}{|c|c|c|c|c|c|}
\hline
&{1}&{2}&{3}&{4}&{5}\\\hline
{1}&0.78&0.18&0.03&0.01&0.00\\\hline
{2}&0.08&0.59&0.29&0.03&0.00\\\hline
{3}&0.01&0.18&0.60&0.20&0.01\\\hline
{4}&0.01&0.03&0.33&0.59&0.04\\\hline
{5}&0.01&0.01&0.12&0.59&0.27\\\hline
\end{tabular} \hspace{.1cm}
\begin{tabular}{|c|c|c|c|c|c|}
\hline
&{1}&{2}&{3}&{4}&{5}\\\hline
{1}&0.92&0.06&0.01&0.00&0.00\\\hline
{2}&0.11&0.65&0.22&0.02&0.00\\\hline
{3}&0.01&0.20&0.68&0.11&0.00\\\hline
{4}&0.00&0.03&0.33&0.58&0.06\\\hline
{5}&0.00&0.01&0.10&0.55&0.34\\\hline
\end{tabular}
\caption{Confusion matrices for Harvard (left), Meridian (middle) and HP-S3 (right).}
\label{tab:confusion}
\end{table}

%

\subsection{Peer Selection}
We demonstrate how peer selection can benefit from network performance prediction, based on ratings of $[1,5]$ in this paper, based on binary classes of ``good'' and ``bad'' in \cite{liao_conext2011} and based on metric values in \cite{liao_dmf2012}. To this end, we let each node randomly select a set of peers from all connected nodes. Each node then chooses a peer from its peer set, and the optimality of the peer selection is calculated by the stretch \cite{Zhang06impactof}, defined as
$$s_i=\frac{x_{i\bullet}}{x_{i\circ}},$$
where $\bullet$ is the id of the selected peer, $\circ$ is that of the true best-performing peer in node $i$'s peer set and $x_{i*}$ is the measured value of some metric. $s_i$ is larger than $1$ for RTT and smaller than $1$ for ABW. The closer $s_i$ is to $1$, the better.

\begin{figure}[t]
\centering
  \begin{tabular}{@{}c@{\hspace{1mm}}c@{\hspace{1mm}}c@{}}
     Harvard & Meridian & HP-S3 \\
\includegraphics[width=0.33\columnwidth]{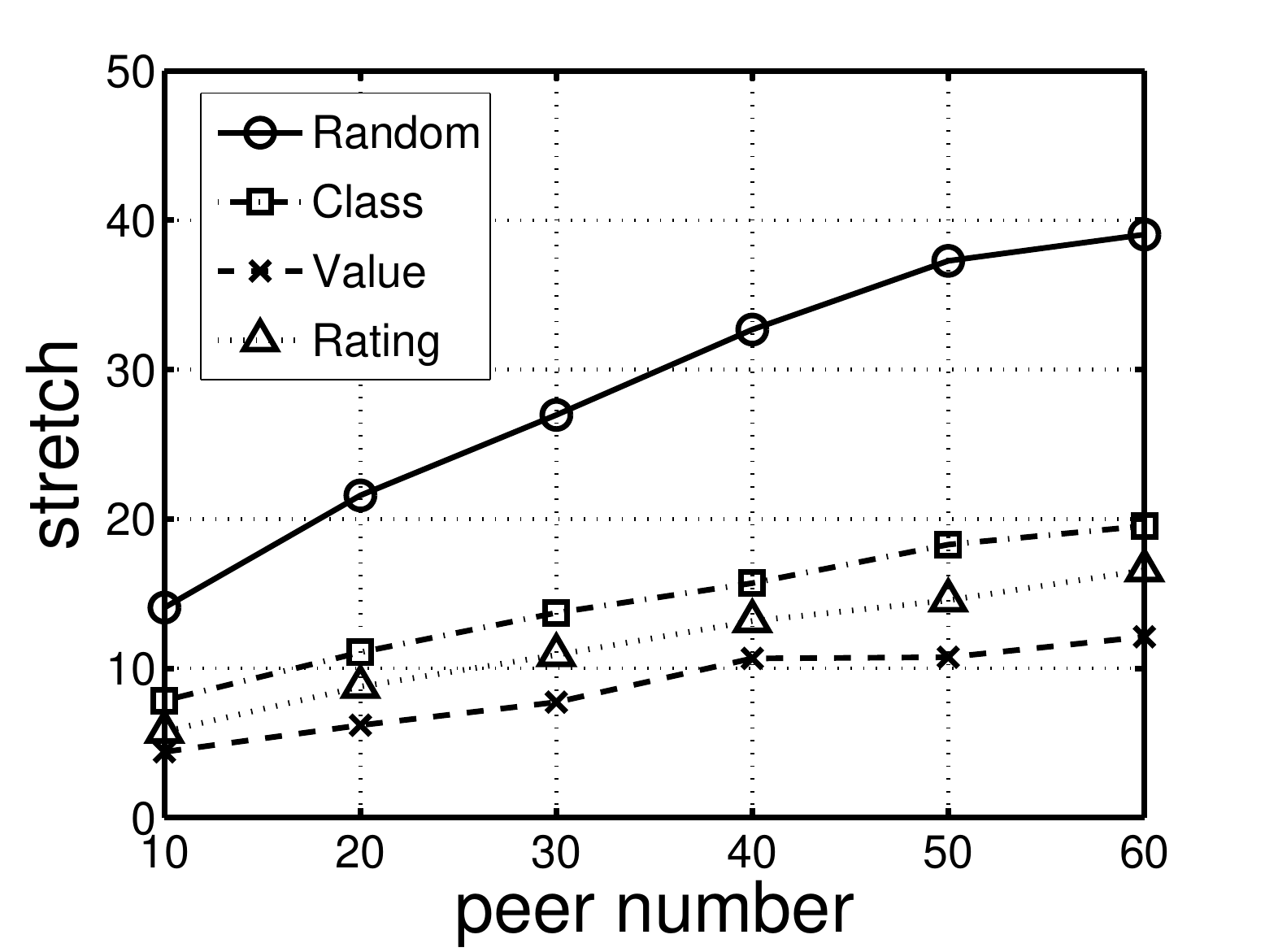}
\label{fig:harvard_peerselection} &
\includegraphics[width=0.33\columnwidth]{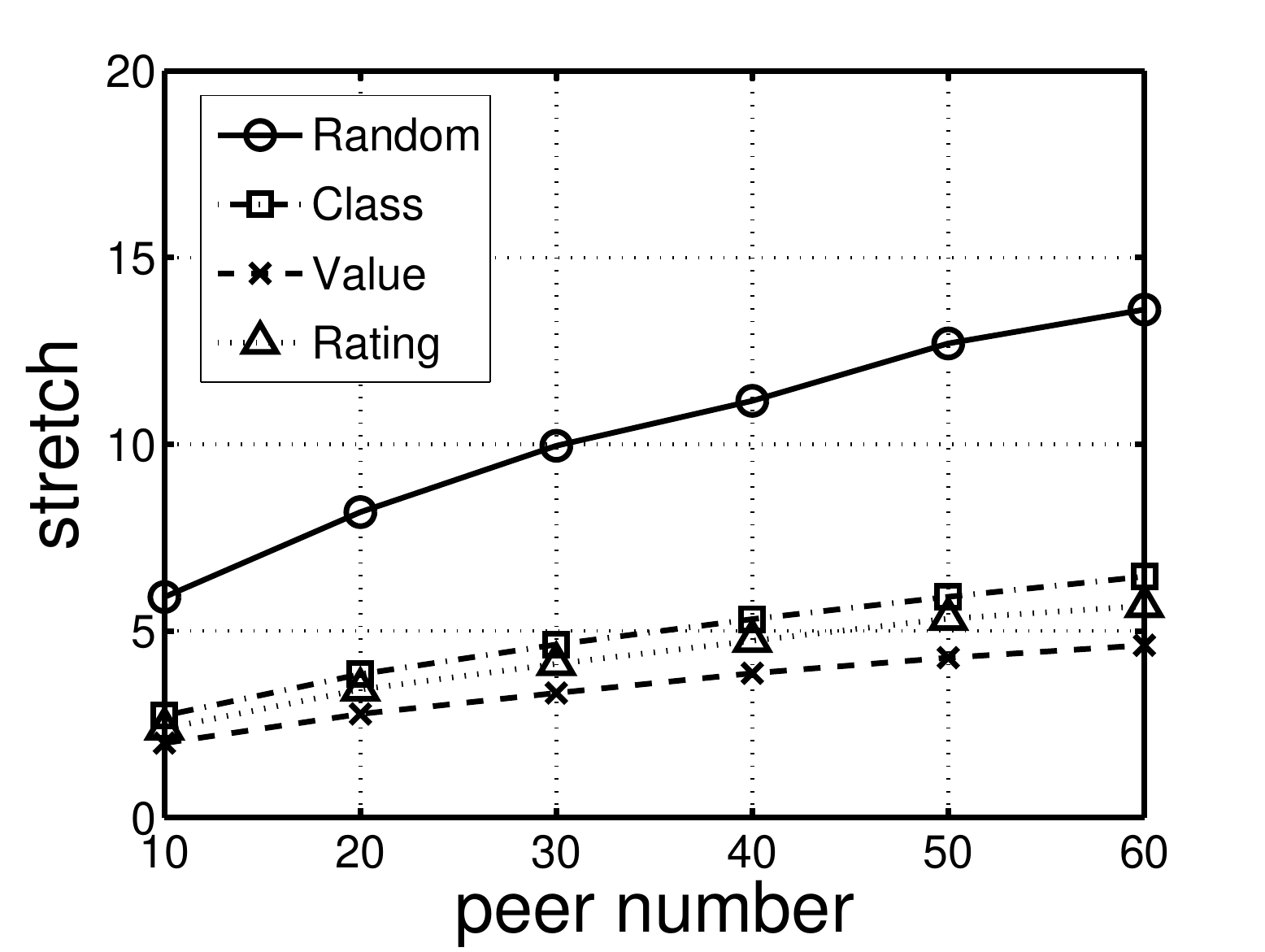}
\label{fig:meridian_peerselection} &
\includegraphics[width=0.33\columnwidth]{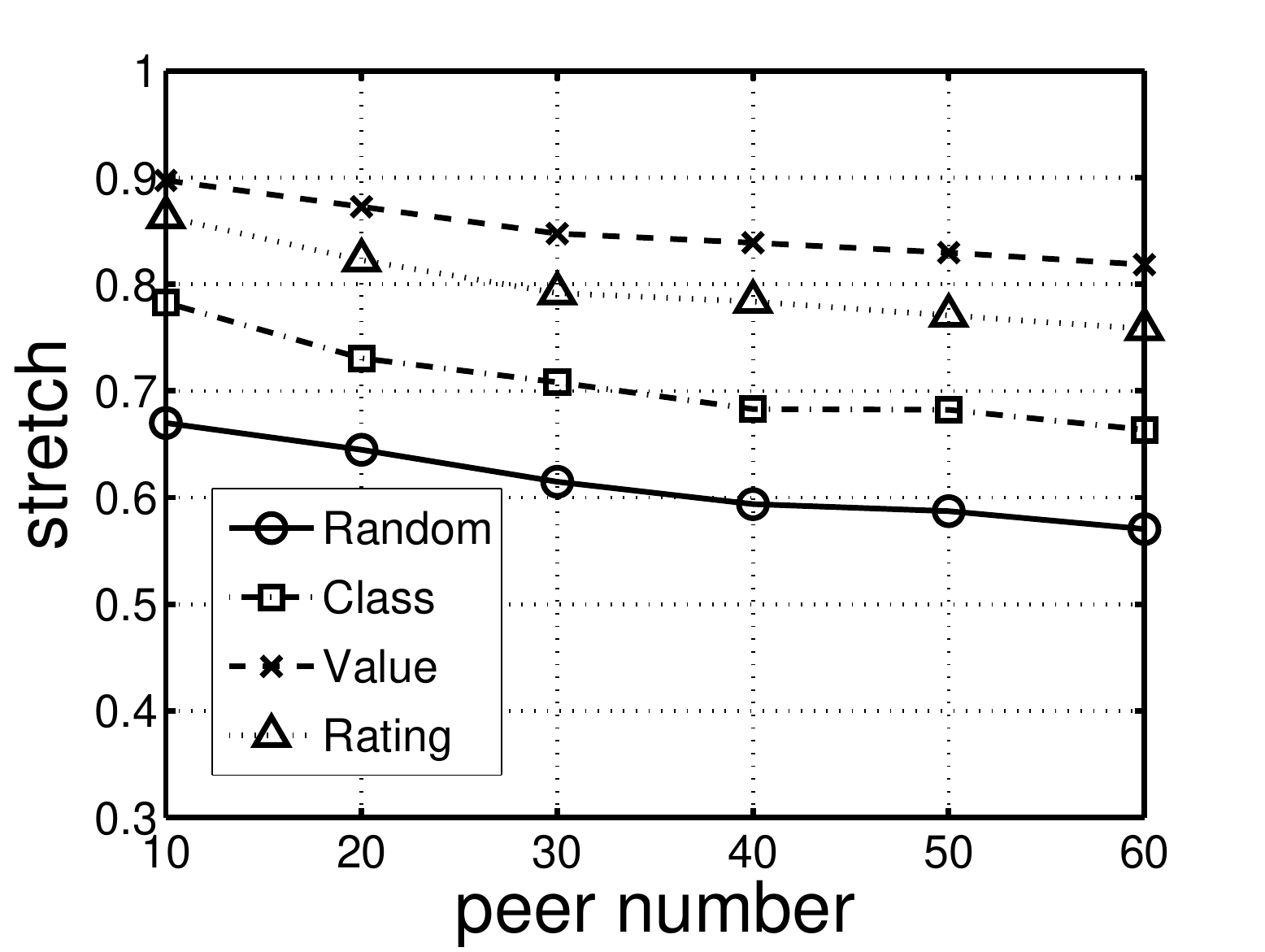}
\label{fig:hp_peerselection}\\
   \end{tabular}
\caption{Peer selection by varying the number of peers in the peer set of each node. Recall that the stretch is larger than $1$ for RTT and smaller than $1$ for ABW. The closer it is to $1$, the better.}
\label{fig:peerselection}
\end{figure}

Figure~\ref{fig:peerselection} shows the stretch of peer selection achieved based on value-based prediction, class-based prediction and our rating-based prediction. Random peer selection is used as a baseline method for comparison. It can be seen that on the optimality, value-based prediction performs the best and the performance by rating-based prediction is better than that of class-based prediction. 
This shows that the rating information is a good comprise between metric values and binary classes. On the one hand, ratings are more informative than binary classes and allow to find better-performing paths. On the other hand, ratings are qualitative and thus require less measurement costs than metric values.

\section{Conclusions}
\label{sec:conclusion}

This paper addresses the scalable acquisition of end-to-end network performance by network inference based on performance ratings. We investigated different matrix factorization models used in Recommender systems, particularly the solution that won the Netflix prize. By taking into account the accuracy and the practicality, the simple regularized matrix factorization was adopted in the inference system. Experiments on peer selection demonstrate the benefit of network inference based on ratings to Internet applications.

\bibliographystyle{splncs}
\bibliography{../references/location.bib}

\end{document}